\documentclass[twocolumn,aps,preprintnumbers,amsmath,amssymb,longbibliography]{revtex4-1}
\usepackage{graphicx}
\usepackage{dcolumn}
\usepackage{bm}
\usepackage{color}
\usepackage{adjustbox}
\usepackage[caption=false]{subfig}
\usepackage{nicefrac}
\usepackage{floatrow}
\usepackage{braket}
\usepackage{tabularx}
\usepackage{soul}
\usepackage{dcolumn}

\begin{document}

\title{Feshbach resonances in $p$-wave three-body recombination within Fermi-Fermi mixtures of open-shell $^6$Li and closed-shell $^{173}$Yb atoms}

\author{Alaina Green$^{1}$}\email{agreen13@uw.edu}
\author{Hui Li$^2$}
\author{Jun Hui See Toh$^1$} 
\author{Xinxin Tang$^1$} 
\author{Katherine McCormick$^1$}
\author{Ming Li$^2$}
\author{Eite Tiesinga$^3$}
\author{Svetlana Kotochigova$^2$}
\author{Subhadeep Gupta$^1$\\
$^1${\em Department of Physics, University of Washington, Seattle WA 98195, USA}\\
$^2${\em Department of Physics, Temple University, Philadelphia, Pennsylvania 19122, USA}\\
$^3${\em Joint Quantum Institute and Joint Center for Quantum Information and Computer Science, 
National Institute of Standards and Technology and University of Maryland, Gaithersburg, Maryland 20899, USA} }

\date{\today}

\begin{abstract}
We report on observations and modeling of interspecies magnetic Feshbach resonances in dilute ultracold mixtures of open-shell alkali-metal $^6$Li and closed-shell $^{173}$Yb atoms with temperatures just above quantum degeneracy for both fermionic species. Resonances are located by detecting magnetic-field-dependent atom loss due to three-body recombination. We resolve closely-located resonances that originate from a weak separation-dependent hyperfine coupling between the electronic spin of $^6$Li  and the nuclear spin of $^{173}$Yb, and confirm their magnetic field spacing by {\it ab initio} electronic-structure calculations. Through quantitative comparisons of theoretical atom-loss profiles and experimental data at various temperatures between 1\,$\mu$K and 20\,$\mu$K, we show that three-body recombination in fermionic mixtures has a $p$-wave Wigner threshold behavior leading to characteristic asymmetric loss profiles. Such resonances can be applied towards the formation of ultracold doublet ground-state molecules and quantum simulation of superfluid $p$-wave pairing.
\end{abstract}

\maketitle

\section{Introduction} 

Magnetic Feshbach resonances (MFRs) are valuable tools in ultracold bosonic and fermionic atomic gases, providing access to tunable interactions between atoms~\cite{chin10,Kotochigova2014}. First observed two decades ago~\cite{inou98,cour98}, they are now routinely used in few- and many-body physics. For example, they are used in the creation of ultracold molecules~\cite{ni08} and in studies of three-body physics~\cite{krae06}. Resonantly-enhanced three-body recombination leads to atom loss from an ultracold sample as well as the formation of tri-atomic Efimov states~\cite{braa06,Naidon2017}. These processes have been studied around weak Feshbach resonances~\cite{DPetrov2004,BoGao2018} and in gases of polar molecules~\cite{Ticknor2010}, while their collision-energy dependence has been examined for bosons~\cite{wang11}. Finally, Feshbach resonances are used to elucidate collective phenomena in Bose-Einstein condensates~\cite{bloc08} and fermionic superfluids~\cite{zwie05}.

Currently, the best studied Feshbach resonances are those between two alkali-metal atoms, each in their open-shell $^2$S ground state. Such resonances are a consequence of atomic hyperfine- and Zeeman-induced interactions between the electron spin singlet and triplet Born-Oppenheimer potentials. The magnetic field dependence of the zero-collision-energy scattering length is given by $a(B)=a_{\mathrm{bg}}\left[1-\Delta/(B-B_{\rm res})\right]$ where $B$ is the magnetic field, $B_{\rm res}$ is the resonance location and $a_{\mathrm{bg}}$ is the background scattering length~\cite{moerdijk1995}. The resonance strength $\Delta$ can be as large as hundreds of Gauss (tens of mT)~\cite{chin10}.

Recently, Feshbach resonances have been observed in mixtures of bosonic alkali-metal and $^1$S alkaline-earth atomic gases (in samples of $^{87}$Rb and $^{87,88}$Sr with nearly equal masses)~\cite{barb18}. Their mixed-species interactions are controlled by a single Born-Oppenheimer potential and, at first glance, no resonances might be expected. However, theoretical predictions have indicated the existence of weak atom-separation-dependent interactions that lead to Feshbach resonances ~\cite{zuch10,brue12,yang2019}. The strongest of these interactions corresponds to a separation-dependent hyperfine coupling between the electron spin of the $^2$S atom and the nuclear spin of the $^1$S atom. The resonances are narrow, with strengths well below one Gauss.

Our research collaboration focuses on ultracold mixtures of $^2$S lithium and alkaline-earth-like $^1$S ytterbium with an extreme mass-imbalance factor of approximately 30. Atomic quantum-degenerate mass-imbalanced mixtures in different isotopic combinations~\cite{hara11,hans11,scha18}, and thus with specific statistical properties, are studied for several reasons. They enable observation of novel fermionic Efimov states in a Fermi-Fermi mixture~\cite{Blume2011,Grimm2019} and may provide a platform for quantum simulation of such phenomena as the Kondo effect~\cite{yaoj19} and induced $p$-wave pairing and superfluidity in Fermi-Bose mixtures~\cite{cara17,kinn18,Blume2011}. Impurity physics in Li and Yb mixtures, where one of the two atomic species is far more prevalent, also has promising basic research implications~\cite{Chevy2006,Cui2010,kinn18}. For example, in a Fermi liquid with impurities near a Feshbach resonance, polarons are an important elementary excitation whose properties will determine the stability of the liquid. Finally, Feshbach resonances among alkali-metal and closed-shell atoms open up the possibility of creating ultracold heteronuclear $^2\Sigma^+$ molecules~\cite{munc11,ciam18,gutt18,barb18,gree19,yang2019}. With the additional degree of freedom from the unpaired electron, such molecules extend the scientific reach of ultracold molecules beyond that of the currently available bi-alkali molecules~\cite{ni08,take14,molo14,park15,Guo16,rvac17} with unique roles in quantum simulation of many-body systems, studies of quantum magnetism, fundamental symmetry tests and ultracold chemistry~\cite{mich06,carr09,gadw16,Makrides2015}.

In this paper we report on first observations and a theoretical study of Feshbach resonances in mixtures of fermionic $^6$Li and fermionic $^{173}$Yb. Using atom-loss spectroscopy as a function of magnetic field with spin-polarized atomic mixtures, we observe a series of interspecies Feshbach resonances corresponding to different magnetic Zeeman states of $^6$Li and $^{173}$Yb. Our observations are in quantitative agreement with our theoretical expectations based on a recently determined Born-Oppenheimer potential~\cite{gree19} and a new electronic valence-bond calculation of separation-dependent hyperfine interactions. We also report on the temperature dependence of the resonance line shapes in conjunction with a model of the resonant interspecies three-body loss processes. As two of the three fermionic atoms in this process are identical, $p$-wave collisions and their corresponding Wigner-threshold behavior play a crucial role, as demonstrated by our analysis. Previously, such temperature analysis has only been performed in single-species bosonic gases~\cite{Laburthe2009,Maier2015,khle19,fouc19} and in mixtures of bosonic $^7$Li and $^{87}$Rb~\cite{Zimmermann2015}.

\begin{figure}
 \includegraphics[width=0.97\columnwidth,trim=0 0 0 0,clip]{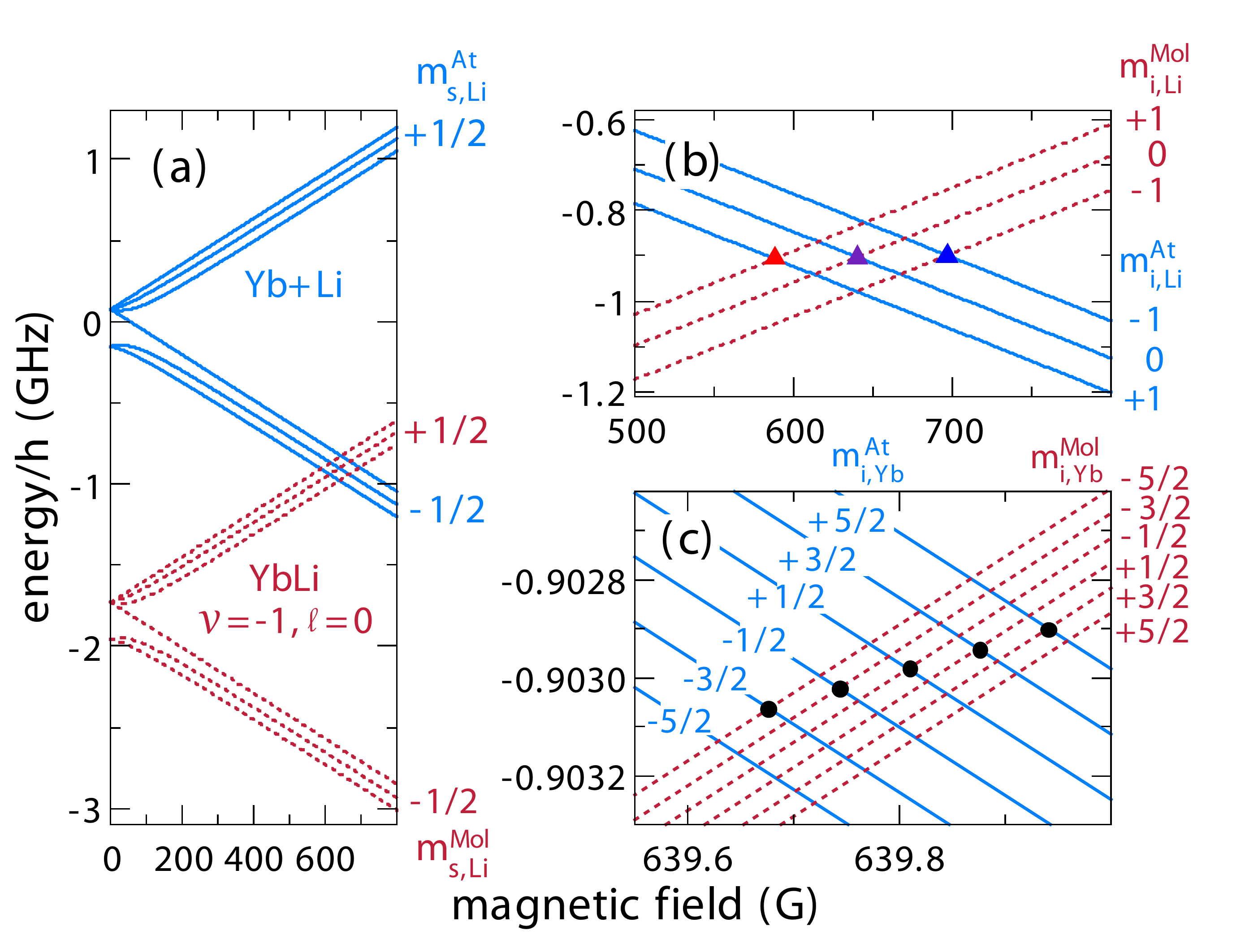}
 \caption{Three views of the hyperfine and Zeeman energies of $^{173}$Yb+$^6$Li zero-energy scattering and bound states as functions of magnetic field. Panel (a) shows the zero-energy atomic energies by solid blue lines and the most weakly-bound $s$-wave bound states by dashed red lines. Panel (b) shows a blowup of the level crossing region in panel (a), while panel (c) shows a further blowup of the crossing region near $640\,$G. Colored symbols in (b) and (c) mark the level crossings for the MFRs observed in this work.}
\label{fig:CrossingOrigin}
\end{figure}

\section{magnetic Feshbach resonances between $^6$Li and $^{173}$Yb \label{sec:origin}} 

Feshbach resonances between ground-state $^6$Li and $^{173}$Yb in a magnetic field $\vec{B}$ are due to the coupling between an electronic Born-Oppenheimer (BO) potential and weak interatomic-separation-dependent hyperfine couplings. The Hamiltonian for this system is $H=H_0+ U_{\rm s\text{-}i}(R)$, where
\begin{eqnarray}
	H_0 = -\frac{\hbar^2}{2\mu}\nabla^2 + V(R) + H_{\rm Li} + H_{\rm Yb}\,,
	\label{eq:h0}
\end{eqnarray}
and $U_{\rm s\text{-}i}(R)$ describes the weak $R$-dependent hyperfine couplings. Here, $R$ is the interatomic separation, ${\mu}$ is the reduced atomic mass, $\hbar=h/(2\pi)$ and $h$ is the Planck constant. The term $V(R)$ represents the ground-state X$^2\Sigma^+$  BO potential.   The last two terms of Eq.~\ref{eq:h0} describe the individual atomic hyperfine and Zeeman Hamiltonians 
\begin{eqnarray}
H_{\rm Li} &=& a_{\rm Li}\,\vec{s}_{\rm Li}\cdot \vec{\imath}_{\rm Li} + ( g^e_{\rm Li}  \vec s_{\rm Li} + g_{\rm Li}^{\rm nuc}\,\vec\imath_{\rm Li})\mu_{\rm B}\cdot {\vec B}\,,   \\
H_{\rm Yb} &=& g_{\rm Yb}^{\rm nuc}\mu_{\rm B}\,\vec\imath_{\rm Yb}\cdot {\vec B},
\label{eq:ath0}
\end{eqnarray}
where the $^6$Li total electron spin is $s_{\rm Li}=\nicefrac{1}{2}$ and its nuclear spin is $i_{\rm Li}=1$. The closed-shell $^{173}$Yb has a nuclear spin $i_{\rm Yb}=\nicefrac{5}{2}$. The $^6$Li hyperfine coupling constant  $a_{\rm Li}$ has units of energy and $g^e_{\rm Li}$, $g_{\rm Li}^{\rm nuc}$, and $g_{\rm Yb}^{\rm nuc}$ are the dimensionless electronic and nuclear $g$ factors of $^6$Li and $^{173}$Yb, respectively. Their values are found in Refs.~\cite{Arimondo1977, olschewski1972}. Finally, $\mu_{\rm B}$ is the Bohr magneton.

The $^6$Li atomic Hamiltonian $H_{\rm Li}$ has magnetic field-dependent eigenstates $|m_{s,\rm Li},m_{i,\rm Li};B \rangle$ labeled by the projection of electron and nuclear spin quantum numbers along $\vec{B}$. We call this the ``high-field'' basis, where the $B$ field label will often be suppressed in states and kets for clarity. Eigenstates of $H_{\rm Yb}$ are $|m_{i,\rm Yb} \rangle$ labeled by the projection of the $^{173}$Yb nuclear spin quantum number along the magnetic field $\vec{B}$. The eigenvalues of  $H_{\rm Li}+H_{\rm Yb}$ as a function of $B$ are shown in Fig.~\ref{fig:CrossingOrigin}. The $^{173}$Yb nuclear Zeeman splittings are only resolved in Fig.~\ref{fig:CrossingOrigin}(c).

The eigenstates of  the  $H_0$ in Eq.~\ref{eq:h0} can be written as
\begin{eqnarray}
|\Psi \rangle= \phi(R) |m_{s,\rm Li},m_{i,\rm Li};B \rangle |m_{i,\rm Yb} \rangle Y_{\ell m_{\ell}}(\hat{R})\,,
\end{eqnarray}
where spherical harmonics $Y_{\ell m_{\ell}}(\hat{R})$ describe the rotation of the two atoms with relative orbital angular momentum $\ell$ and its projection $m_{\ell}$. The radial wave function $\phi(R)$ is either a scattering solution of the BO potential at collision energy $E>0$ and partial wave $\ell$ or a bound state with energy $E_{\nu,\ell}<0$, where $\nu$ and $\ell$ are the vibrational and rotational quantum numbers respectively. Throughout this paper these two types of solutions and spin states are distinguished by superscripts At and Mol where necessary. We use the X$^2\Sigma^+$ BO potential from Ref.~\cite{gree19}, which was obtained by fitting to six experimentally-determined weakly-bound states of the isotopologue $^{174}$Yb$^6$Li. The most weakly-bound state of $^{173}$Yb$^6$Li in this potential has energy $E_{-1,0}/h = -1.8058(34)\,$GHz, where the number in parentheses is the one-standard deviation uncertainty. 

Bound-state energies of $H_0$ are the sum of the $^6$Li and $^{173}$Yb hyperfine and Zeeman energies and $E_{\nu,\ell}$. Binding energies of the hyperfine levels with $\nu=-1$ and $\ell =0$ as functions of $B$ are shown in Figs.~\ref{fig:CrossingOrigin}(a) and (b). Crossings between atomic and molecular states are visible, although the states are not coupled within $H_0$. Once we include the weak $U_{\rm s\text{-}i}(R)$ interactions they change into Feshbach resonances. We postpone the description of this mixing until Sec.~\ref{sec:rdepU} and first describe our experimental procedures to precisely locate the resonances.

\section{Experimental Setup}\label{sec:Setup}

We observe interspecies magnetic Feshbach resonances through  enhanced  atom loss over narrow ranges of magnetic field. Specifically, we measure the remaining fraction of atoms after an optically-trapped spin-polarized mixture is held for a fixed time at constant magnetic field. This atom-loss spectroscopy begins with ultracold atomic samples in a crossed optical dipole trap (ODT) described in earlier work~\cite{Roy16,roy17}. Unpolarized laser-cooled samples of atomic $^{173}$Yb are loaded into the ODT first at a bias magnetic field of $\lesssim 1\,$G. Subsequently, $^6$Li atoms are laser cooled, loaded into the ODT, and optically pumped into the two energetically-lowest hyperfine states by applying light resonant on the $^2{\rm S}_{\nicefrac{1}{2}}\,\rightarrow\,^2{\rm P}_{\nicefrac{3}{2}}$ transition. We then increase the bias magnetic field to $\simeq 500\,$G in order to spectroscopically resolve the two remaining hyperfine states and subsequently remove atoms in the $\ket{m_{s,\rm Li},m_{i,\rm Li}}=\ket{-\nicefrac{1}{2},+1}$ state with an additional resonant light pulse. 

\begin{figure}
	\includegraphics[width=1.0 \columnwidth]{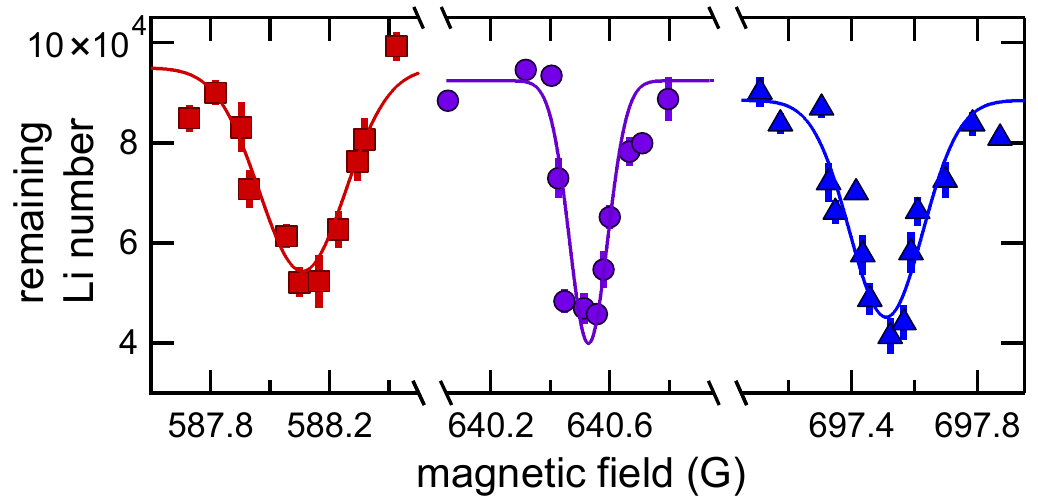}
\caption{Observation of interspecies $^{173}$Yb$^6$Li magnetic Feshbach resonances as functions of magnetic field $B$. Resonances appear as $B$-dependent atom loss. The $^6$Li atom loss is shown after the $2.8\,\mu$K mixture has been held for $3\,$\,s in the dipole trap. Red squares, purple circles, and blue triangles correspond to data for $^{6}$Li in states $|m_{s,\rm Li},m_{i,\rm Li}\rangle=|-\!\nicefrac{1}{2},1\rangle$, $|-\!\nicefrac{1}{2},0\rangle$, and $|-\!\nicefrac{1}{2},-1\rangle$, respectively. The nuclear Zeeman state of $^{173}$Yb is spin-polarized to $|m_{i,\rm Yb}\rangle=|\!+\!\nicefrac{5}{2}\rangle$ in each case. Each point is the average of at least four measurements and the error bars are one-standard-deviation statistical uncertainties. Curves are best-fit Gaussians and only meant as a guide to the eye. Our full line shape analysis is found in Sec.~\ref{sec:temp}.}
	\label{fig:LiSpin}
\end{figure}

To prepare a particular spin-polarized heteronuclear mixture we use the following strategy (additional details can be found in the Supplemetary Material~\cite{supp}). A first stage of evaporative cooling to $5.8\,\mu$K is performed by ODT depth reduction before the $^{173}$Yb sample is partially polarized through optical pumping  into a spin mixture containing a majority of $^{173}$Yb atoms in state $\ket{m_{i,\rm Yb}=m}$ and a minority in a sacrificial state $\ket{m_{i,\rm Yb}=+\nicefrac{5}{2}}$ or $\ket{m_{i,\rm Yb}=-\nicefrac{5}{2}}$. The sacrificial state is retained to increase the efficiency of further evaporative cooling. Subsequently, the sample temperature is either increased or decreased to the desired value by either increasing or decreasing the ODT depth~\footnote{Decreasing the ODT depth leads to a continuation of the evaporative cooling process. Increasing the depth of our ODT also reduces the volume of the gas and hence increases the temperature, with a lower limit set by an unchanged phase space density.}. The $^{173}$Yb atoms in the sacrificial state are then removed with a resonant light pulse,  resulting in a fully polarized $^{173}$Yb sample in state $\ket{m_{i,\rm Yb}=m}$. Finally, the atoms in the $^6$Li sample are transferred to the hyperfine ground state of interest through radiofrequency (RF) adiabatic rapid passage. We have verified from a separate diagnostic that we achieve $>90\%$ spin polarization for each atomic species in the targeted spin state~\cite{supp}. 

Once the desired spin-polarized heteronuclear mixture is prepared, we smoothly ramp the magnetic field to a specific value to perform loss spectroscopy. The spectroscopy phase consists of letting the atoms interact for a fixed hold time at this magnetic field after which we measure the remaining atom number through absorption imaging at $500\,$G. We then repeat the process for many magnetic field values. The magnetic field is generated by a pair of coils in Helmholtz configuration connected to a programmable power supply, and is calibrated through RF spectroscopy on the $^6$Li atomic hyperfine transitions. 

The temperature range explored in this work is between $1\,\mu$K and $20\,\mu$K. The differential gravitational potential in this highly mass-imbalanced system results in a partial separation of the two species at the lowest temperatures, causing a lengthening of interspecies thermalization time below $\simeq 2\,\mu$K. In this work, the lowest $^{173}$Yb ($^6$Li) temperature at the beginning of the loss spectroscopy phase is $1.0\,\mu$K ($1.8\,\mu$K). This corresponds to $T/T_F=3.3$ (0.57), where $T_F$ is the Fermi temperature for each species. Under these conditions, the measured $^{173}$Yb ($^6$Li) atom number at the beginning of the spectroscopy phase is $1.0\times 10^5$ ($1.3 \times 10^5$) with corresponding peak density of $2.6 \times 10^{12}$\,cm$^{-3}$ ($6.1 \times 10^{12}$\,cm$^{-3}$). Here and elsewhere in this paper, the uncertainties in temperature, atom number and density are $10\%$, $10\%$ and $18\%$ respectively, mainly stemming from uncertainties in the imaging system. For higher temperatures, the two species are in thermal equilibrium with each other at the beginning of the loss spectroscopy phase.

\section{Observation of Magnetic Feshbach Resonances}\label{sec:Obs}

For the magnetic field range investigated in this work three $^6$Li ground hyperfine states exhibit interspecies Feshbach resonances with $^{173}$Yb. These are  $|m_{s,\rm Li},m_{i,\rm Li}\rangle=\ket{-\nicefrac{1}{2},+1}$, $\ket{-\nicefrac{1}{2},0}$, and  $\ket{-\nicefrac{1}{2},-1}$. Fig.~\ref{fig:LiSpin} shows the experimental observation of their interspecies Feshbach resonances when the $^{173}$Yb is prepared in $|m_{i,\rm Yb}\rangle=|+\nicefrac{5}{2}\rangle$. We have confirmed that the three loss features in Fig.~\ref{fig:LiSpin} correspond to interspecies Feshbach resonances by repeating the spectroscopy phase with only $^6$Li atoms. No atom loss features were then observed. 

\begin{figure}
	\includegraphics[width=0.95\columnwidth]{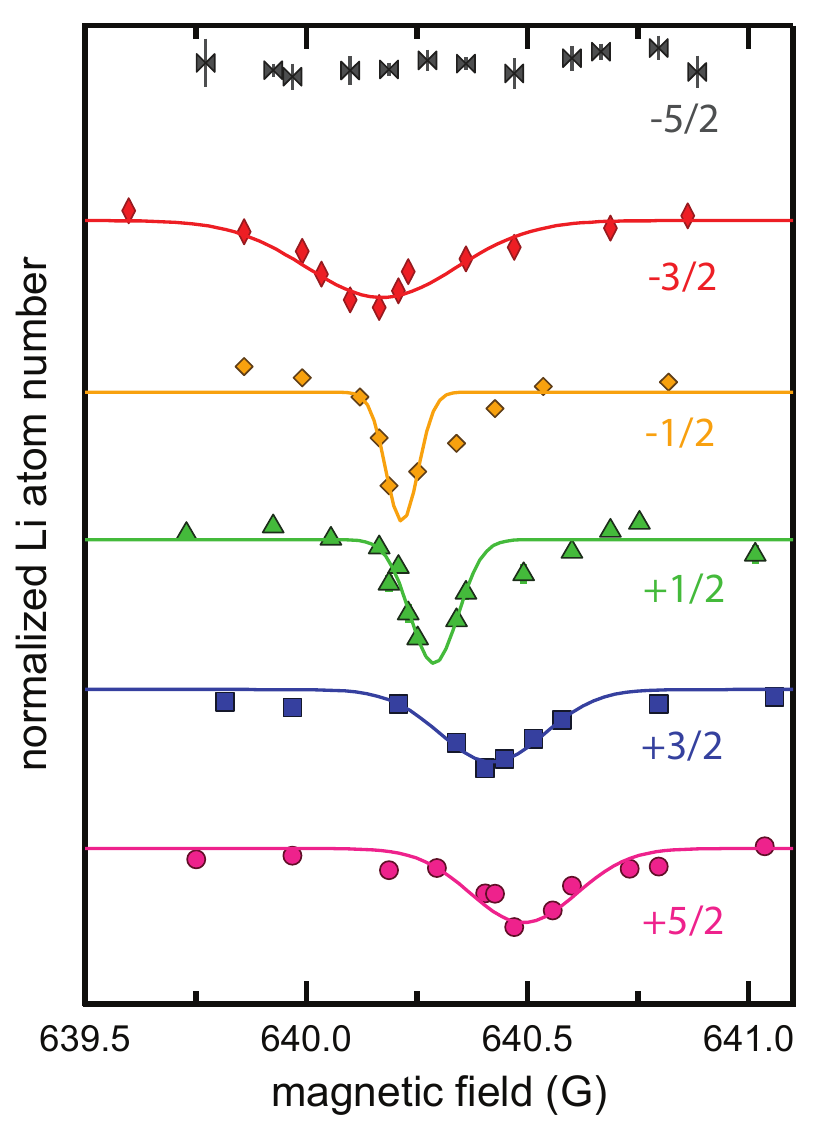}
\caption{Dependence of magnetic Feshbach resonances on the $^{173}$Yb nuclear Zeeman state in a $^6$Li$^{173}$Yb mixture seen in the remaining $^6$Li atom number as functions of $B$ near the $640\,$G Feshbach resonance. $^6$Li is prepared in hyperfine state $|m_{s,\rm Li},m_{i,\rm Li}\rangle=|-\nicefrac{1}{2},0\rangle$ and the $^{173}$Yb sample is spin polarized in different nuclear Zeeman states. The data is normalized by the remaining $^6$Li atom number away from the resonance and offset for clarity. From top to bottom black, red, yellow, green, blue, and pink markers correspond to data with $^{173}$Yb prepared in nuclear spin state $m_{i,\mathrm{Yb}}=-\nicefrac{5}{2}, -\nicefrac{3}{2},-\nicefrac{1}{2},+\nicefrac{1}{2}, +\nicefrac{3}{2}$, and $+\nicefrac{5}{2}$, respectively. No resonance exists for $m_{i,\mathrm{Yb}}=-\nicefrac{5}{2}$. The temperature is $1.8\,\mu$K ($1.0\,\mu K$) for Li (Yb) and the hold time is $1.5\,$s. Curves are best-fit Gaussians. Our full line shape analysis is found in Sec.~\ref{sec:temp}.}
	\label{fig:YbSpin}
\end{figure}

To investigate the effect of the $^{173}$Yb nuclear spin on the MFRs, we repeated our trap-loss spectroscopy for each of the six $m_{i,\mathrm{Yb}}$ states, preparing the $^6$Li sample in $|m_{s,\rm Li},m_{i,\rm Li}\rangle=\ket{-\nicefrac{1}{2},0}$ for all cases. The results are shown in Fig.~\ref{fig:YbSpin}. The absence of a MFR for $m_{i,\mathrm{Yb}}=-\nicefrac{5}{2}$ is expected for reasons outlined in Sec.~\ref{sec:rdepU}. 

The experimental value of $B_{\mathrm{res}}$ for each MFR is determined as the center value of a Gaussian fit to our lowest temperature data~\footnote{We have verified that the shifts between adjacent resonances determined by Gaussian fits agree with the results determined by the full line shape analysis in Sec.~\ref{sec:temp} within the experimental error bars.}. The locations of all observed resonances with spin-polarized heteronuclear mixtures are listed in Table~\ref{tab:FRcomp} and are consistent with the predictions for Feshbach resonance locations due to the least bound state, i.e. $\nu=-1, \ell=0$, of the $^2\Sigma^+$ BO potential shown in Fig. \ref{fig:CrossingOrigin}. We present a detailed comparison of our theoretical analysis and experimental observations in Sec.~\ref{sec:comp}.

The locations of the five resonances in Fig.~\ref{fig:YbSpin} are also indicated in Fig.~\ref{fig:wfhyp}(a). From these locations we derive the magnetic moment of the $^{173}$Yb nucleus in the $\nu=-1, \ell=0$ $^{173}$Yb$^6$Li molecule and note that its sign is opposite that of the free $^{173}$Yb atom. In Sec. \ref{sec:rdepU} and \ref{sec:comp} we show that this sign change originates from the $R$-dependent $U_{\rm s-i}(R)$ hyperfine coupling.

\section{Separation-dependent hyperfine interactions} \label{sec:rdepU}

\begin{figure}
	\includegraphics[scale=0.4,trim=0 0 0 0,clip]{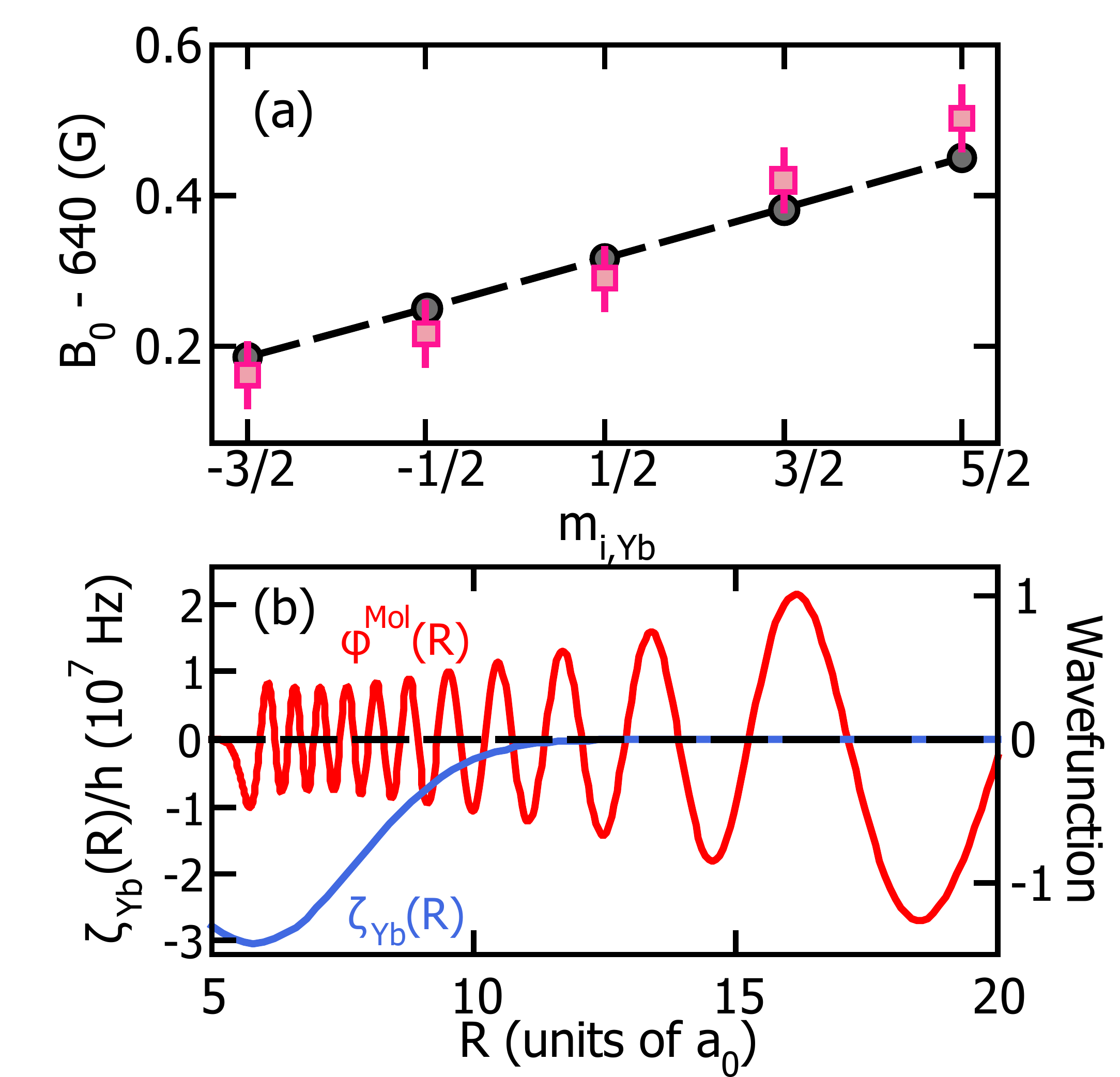}
	\caption{(a) The  observed (pink squares) and theoretically predicted (black circles) resonance locations near $B=640$\,G as functions of $m_{i,\rm Yb}$. The $^6$Li atom is in state $|m_{s, \rm Li},\ m_{i,\rm Li}\rangle=|-\nicefrac{1}{2},0\rangle$. The theoretical values have been uniformly shifted from those shown in Fig.~\ref{fig:CrossingOrigin}(c) as explained in the text. The dashed line is given by Eq.~\ref{eq:de1}. (b) The $^{173}$Yb$^6$Li hyperfine coupling coefficient $\zeta_{\rm Yb}(R)$ (solid blue line) and the radial wave function $\phi^{\rm Mol}(R)$ of the most weakly-bound state of the $^2\Sigma^+$ potential (solid red line) as functions of separation $R$. Here, $a_0=0.05292$ nm is the Bohr radius. }
	\label{fig:wfhyp}
\end{figure}

We now define the weak interaction $U_{ \rm s\text{-}i }(R)$ that leads to coupling between eigenstates of $H_0$ and hence the Feshbach resonances. The interaction describes the effects of the modified electron spin densities at the nuclear positions of $^6$Li and $^{173}$Yb when the atoms are in close proximity. For our experiment the relevant coupling is 
\begin{eqnarray}
U_{ \rm s\text{-}i }(R) = \zeta_{\rm Yb}(R)\, \vec{s}_{\rm Li}\cdot \vec{\imath}_{\rm Yb}\,,
\label{eq:si}
\end{eqnarray}
where the hyperfine coupling coefficient $\zeta_{\rm Yb}(R)$ is obtained from an all-electron {\it ab initio} calculation based on the non-relativistic configuration interaction valence-bond (CI-VB) method~\cite{Tupitsyn1998,Kotochigova1998,Kotochigova2005}.
Fig.~\ref{fig:wfhyp}(b) shows $\zeta_{\rm Yb}(R)$ together with the real-valued radial wave function of the most weakly-bound state of the $^2\Sigma^+$ BO potential as a function of interatomic separation. The $\zeta_{\rm Yb}(R)$ is on the order of $a_{\rm Li}$ near the inner point, $R \approx 6 a_0$, of the vibrational bound state wave function and then approaches zero rapidly when $R\to\infty$. Other weaker coupling terms~\cite{zuch10,brue12,yang2019} are discussed in the Supplementary Material~\cite{supp}. 

The weak $U_{ \rm s\text{-}i }(R)$ changes the crossings between the atomic and molecular levels in Fig.~\ref{fig:CrossingOrigin} into resonances. For this interaction to lead to a resonance the sum $m_{s,\rm Li}+m_{i,\rm Yb}$ must be the same for the scattering and bound states. In particular, since the MFRs we investigate satisfy $m^{\rm At}_{s,\rm Li}\,\left(m^{\rm Mol}_{s,\rm Li}\right) = -\nicefrac{1}{2}\,\left(+\nicefrac{1}{2}\right)$ for the scattering (bound) states, no MFR is expected for $m^{\rm At}_{i,\rm Yb} = -\nicefrac{5}{2}$.  Additionally, the projection $M_{\rm tot} = m_{s,\rm Li}+m_{i,\rm Li}+m_{i,\rm Yb}$ is always conserved.

The weak $U_{ \rm s\text{-}i}(R)$ also modifies the energies of the hyperfine and Zeeman states of the ${\nu=-1}$, $\ell =0$ bound state. The size of these energy shifts can only be observed over magnetic field intervals of less than 1\,G as shown in Fig.~\ref{fig:CrossingOrigin}(c) for resonances around 640\,G. These resonances are labeled by the $^6$Li state $|m^{\rm At}_{s,\rm Li},m^{\rm At}_{i,\rm Li};B \rangle = |-\nicefrac{1}{2},0;B \rangle$ and $^{173}$Yb Zeeman states $|m^{\rm At}_{i,\rm Yb} \rangle$ for the scattering states and $|m^{\rm Mol}_{s,\rm Li},m^{\rm Mol}_{i,\rm Li};B \rangle = |\nicefrac{1}{2},0;B \rangle$ and $|m^{\rm Mol}_{i,\rm Yb} \rangle$ for the molecular states. 

To get an intuitive understanding of the molecular level splitting, we perturbatively study the energies of these six bound states. The states of different $m_{i,\rm Yb}$ are split by the nuclear Zeeman interaction of $^{173}$Yb,  
\begin{eqnarray}
\Delta E_0 = m_{i,\rm Yb}\, g^{\rm nuc}_{\rm Yb} \mu_{\rm B}B\,,
\label{eq:de0}
\end{eqnarray}
and a contribution from the diagonal matrix elements of Eq.~\ref{eq:si}
\begin{eqnarray}
\Delta E_1 &=&  \langle \Psi^{\rm Mol} |\zeta_{\rm Yb}(R)\, \vec{s}_{\rm Li}\cdot \vec{\imath}_{\rm Yb}| \Psi^{\rm Mol} \rangle \nonumber \\
& \simeq & \frac{1}{2} m_{i,\rm Yb} \int_{0}^{\infty} dR\,\phi^{\rm Mol}(R) \zeta_{\rm Yb}(R) \phi^{\rm Mol}(R)\,,
\label{eq:de1}
\end{eqnarray}
where for the last equality we used the fact that the magnetic field is large. The radial integral over $\zeta_{\rm Yb}(R)$ is $-0.369 \times h$\,MHz for our CI-VB values. Both Eqs.~\ref{eq:de0} and \ref{eq:de1} are proportional to $m_{i,\rm Yb}$ and the two contributions have opposite signs. For $B=640$\,G, $|\Delta E_1 |$ is about $30\%$ larger than $\Delta E_0$, resulting in an overall change in the sign of the level shifts as compared to shifts of the free-atom state. 

Figure \ref{fig:CrossingOrigin}(c) shows the theoretical energies of the atomic levels crossing with the molecular bound states including the corrections $\Delta E_0$ and $\Delta E_1$. The energies of scattering states are not affected by $U_{ \rm s\text{-}i}(R)$. Crossings with markers in Fig.~\ref{fig:CrossingOrigin}(c) correspond to resonances satisfying the selection rules of $U_{ \rm s\text{-}i}(R)$. Without the correction of Eq.~\ref{eq:de1} in molecular state energies, all crossings would occur at the same magnetic field~\cite{yang2019}.

We use our coupled-channels code to determine the strength and resonance locations of the MFRs. Specifically, we compute the zero-energy $s$-wave scattering length $a(B)$ as a function of magnetic field for all relevant scattering channels $|m_{s, \rm Li}^{\rm At},\ m_{i,\rm Li}^{\rm At}\rangle$. Near each resonance, we fit to $a(B)=a_{\rm bg}[1-\Delta/(B-B_{\rm res})]$ as defined in the introduction \footnote{In practice, we use collision energy $E/k_B=$ 100 nK and have observed that inelastic losses are negligibly small. Here, $k_B$ is the Boltzmann constant.}.

\section{Comparison between theory and experiment}\label{sec:comp}

Table~\ref{tab:FRcomp} lists our observed resonance locations as well as the corresponding theoretical predictions of resonance locations, strengths and other properties based on the BO potential that gives $E_{-1,0}/h = -1.8058$ GHz for the most weakly-bound state of $^{173}$Yb$^6$Li. Near 640\,G, we experimentally located the resonances for all nuclear Zeeman states of $^{173}$Yb. No resonance exists for $m_{i,\rm Yb}=-\nicefrac{5}{2}$. Similar families of resonances occur near 588 and 697\,G (see Supplementary Material~\cite{supp}). We note that the observed and theoretical locations are consistent, as the 3.4\,MHz uncertainty in the binding energy of the $\nu=-1$ and $\ell =0$ state and the approximately 2$\mu_B$ magnetic moment difference between the bound and scattering states leads to a 1.3\,G uncertainty in the theoretical resonance location. All observed locations occur at a larger magnetic field value than those of the theoretical predictions, indicating that the binding energy $\left|E_{-1,0}\right|$ is slightly underestimated. 

Figure~\ref{fig:wfhyp}(a) compares the experimental and theoretical resonance locations near 640\,G as functions of $m_{i,\rm Yb}$. For better visual comparison, the theoretical values have been uniformly shifted up by 0.58\,G, a value within the 1.3\,G uncertainty. The theoretical locations have a  linear dependence on $m_{i,\rm Yb}$ with a slope solely determined by Eq.~\ref{eq:de1}. The experimental values are consistent with this linear dependence. This gives us confidence in our CI-VB calculation of the $R$-dependent $\zeta_{\rm Yb}(R)$.  

\begin{table}
	\caption{Observed $^{173}$Yb$^6$Li Feshbach resonance locations and corresponding theoretical predictions and assignments based on $s$-wave coupled-channels calculations. The first two columns describe the quantum numbers of the scattering states and the projection of the total angular momentum, respectively. The third and fourth columns give the observed and predicted resonance locations. Finally, the last column gives the resonance strength from the coupled-channels calculations. The error in the observed locations is the one-standard-deviation uncertainty from the quadrature sum of the statistical error in the Gaussian fit and the systematic error in the field calibration. The theoretical locations of the resonances have a 1.3\,G one-standard-deviation uncertainty due to the uncertainty of the binding energy of the most weakly-bound state. The differences between neighboring resonances are not affected by this uncertainty.}
	\begin{ruledtabular}
	\begin{tabular}{ccccc}
		\noalign{\smallskip}		
		${|m_{s, \rm Li}^{\rm At},\ m_{i,\rm Li}^{\rm At}\rangle\!+\!|m_{i,\rm Yb}^{\rm At}\rangle}$ & ${\rm M_{tot}}$ & $B^{\rm Exp}_{\rm res}$ (G) & $B^{\rm The}_{\rm res}$ (G) & $\Delta$ ($\mu$G) \\
		\noalign{\smallskip}
		\hline
		\noalign{\smallskip}
		$|-\nicefrac{1}{2}, 1\rangle$ + $|+\nicefrac{5}{2}\rangle$ &  3 & 588.126(41) & 587.803 & 27.6  \\
		\hline
		$|-\nicefrac{1}{2}, 0\rangle$ + $|-\nicefrac{3}{2}\rangle$ & -2 & 640.161(40) & 639.605 & 27.6  \\
		$|-\nicefrac{1}{2}, 0\rangle$ + $|-\nicefrac{1}{2}\rangle$ & -1 & 640.216(41) & 639.670 & 44.2  \\
		$|-\nicefrac{1}{2}, 0\rangle$ + $|+\nicefrac{1}{2}\rangle$ &  0 & 640.289(40) & 639.736 & 49.7  \\
		$|-\nicefrac{1}{2}, 0\rangle$ + $|+\nicefrac{3}{2}\rangle$ &  1 & 640.420(40) & 639.802 & 44.2  \\
		$|-\nicefrac{1}{2}, 0\rangle$ + $|+\nicefrac{5}{2}\rangle$ &  2 & 640.502(44) & 639.870 & 27.6  \\
		\hline
		$|-\nicefrac{1}{2}, -1\rangle$ +$|+\nicefrac{5}{2}\rangle$ &  1 & 697.523(40) & 696.545 & 27.6  \\
	\end{tabular}
\end{ruledtabular}
\label{tab:FRcomp}
\end{table} 

\section{Fermionic features in resonant atom-loss line shapes} \label{sec:temp}

Our atom loss measurements also confirm the fermionic statistical properties of our mixture. The requirement of an anti-symmetric scattering wavefunction under interchange of identical fermions leads to a three-body loss rate coefficient that has a ``$p$-wave Wigner threshold character''. Specifically, we will show that the locations of the maxima of the three-body loss rates as functions of $B$ shift linearly with increasing temperature in a way that can only be explained by the fermionic nature of the scattering atoms. Our data are also consistent with the observation that the maximum event rate coefficient must also be independent of temperature.

We start by noting that atom loss from our mixture at temperature $T$ is described by the two-coupled equations
\begin{subequations}
\begin{eqnarray}
\frac{dN_{\rm Li} }{dt} & = & -\Gamma_{\rm Li} N_{\rm Li} - 2  \gamma_1
N^2_{\rm Li} \,N_{\rm Yb} - \gamma_2 N_{\rm Li} \,N^2_{\rm Yb}\,, \\
\frac{dN_{\rm Yb} }{dt} & = & -\Gamma_{\rm Yb} N_{\rm Yb} - \gamma_1
N^2_{\rm Li} \,N_{\rm Yb} -2 \gamma_2 N_{\rm Li} \,N^2_{\rm Yb}\,,
\end{eqnarray}
\label{eq:ant}
\end{subequations}
\!\!\!where atom numbers $N_{a}$ for $a$\,=\,$\mathrm{Li}$ or Yb are time dependent. Rates $\Gamma_a$ describe one-body background-collision-induced losses, while event rates $\gamma_1$ and $\gamma_2$ describe the three-body recombination processes starting from $^6$Li+$^6$Li+$^{173}$Yb and $^6$Li+$^{173}$Yb+$^{173}$Yb collisions, respectively. For both processes two of the three atoms are identical fermions. Finally, $\gamma_i= K_i(B,T)/V_i$ with $i=$1 or 2 and temperature-dependent hypervolumes \[
\frac{1}{V_1} = \int d^3x \,\rho^2_{\rm Li}(\vec{x}) \rho_{\rm Yb}(\vec{x})
\ {\rm and}\ \frac{1}{V_2} = \int d^3x \,\rho_{\rm Li}(\vec{x}) \rho^2_{\rm Yb}(\vec{x})\,, \] where the time-independent $\rho_a(\vec x)$ are unit-normalized spatial density profiles. The event rate coefficients $K_i(B,T)$ will be discussed below. 

Several assumptions have gone into deriving Eqs.~\ref{eq:ant} from two coupled Boltzmann equations for the single-particle phase space densities $f_a(\vec{x},\vec{p},t)$~\cite{luiten1996kinetic,anderlini2006thermalization} with momentum $\vec{p}$. We assume that both fermionic $^6$Li and $^{173}$Yb gases are in thermal equilibrium at a temperature above degeneracy and $f_a(\vec{x},\vec{p},t) \propto N_a(t)\rho_a(\vec x)\exp[-p^2/(2m_ak_BT)]$, where $m_a$ is the mass of atom $a$ and $k_B$ is the Boltzmann constant. This is justified as the mean time between thermalizing elastic Yb+Li collisions is much smaller than the time scales of atom loss due to three-body recombination~\cite{ivan11,hara11,gree19}. Even though the two species are held in the same dipole trap, their spatial density profiles $\rho_a(\vec x)$ are distinct as their dynamic polarizabilities and gravitational potentials are different. In this section, however, the small differences in temperature and spatial density profiles between the $^6$Li and $^{173}$Yb gases will be ignored. In fact, differences are only significant for our smallest measured temperature, where quantum degeneracy is almost reached and the thermalization times are longest. Finally, losses from two-body Li+Yb collisions are negligible as confirmed by our coupled-channels calculations. Other two- and three-body losses are suppressed by the fermionic nature of the $^6$Li and $^{173}$Yb atoms.  

\begin{figure*}
	\includegraphics[scale=0.30,trim=0 0 0 0,clip]{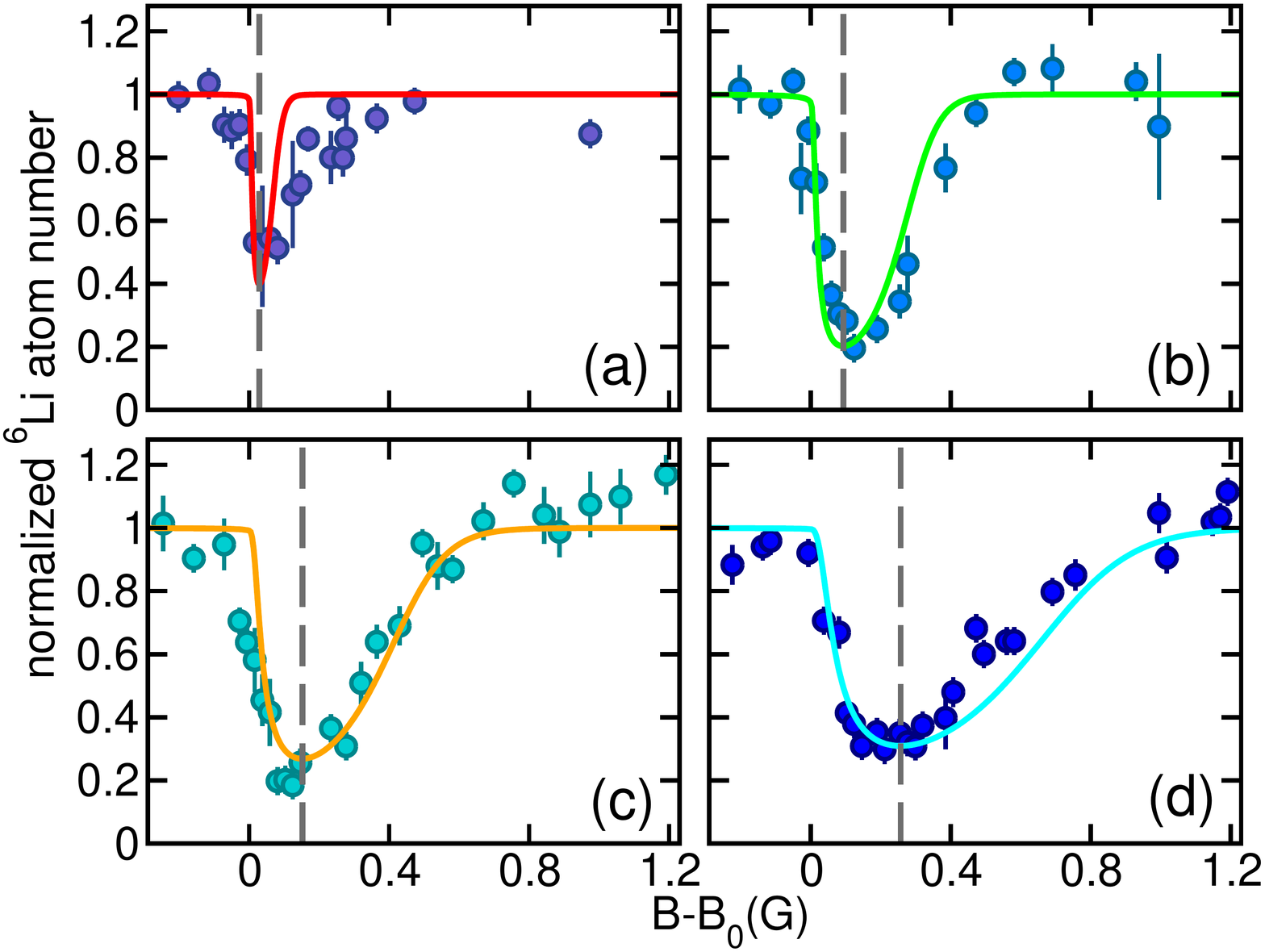}
	\includegraphics[scale=0.30,trim=0 0 0 0,clip]{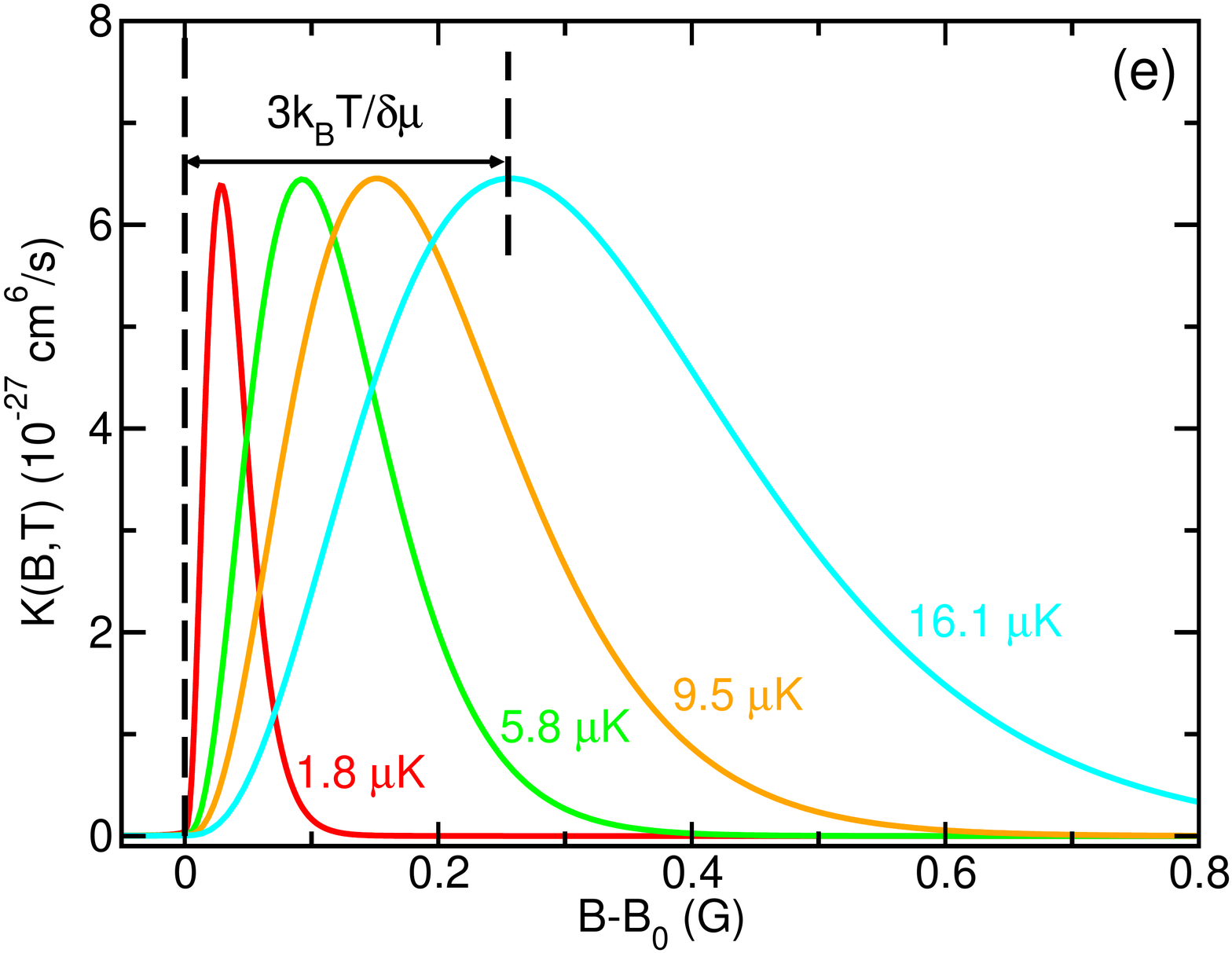}
  \caption{Fermionic behavior of three-body recombination processes in $^6$Li and $^{173}$Yb mixtures found in atom-loss spectra as functions of magnetic field. $^6$Li is prepared in the $|m_{s,\rm Li},m_{i,\rm Li}\rangle=|-\nicefrac{1}{2},0\rangle$ state and $^{173}$Yb in the $|m_{i,\rm Yb}\rangle=|+\nicefrac{5}{2}\rangle$ state. The hold time is 4\,s.  Panels \{(a),(b),(c),(d)\} correspond to the remaining $^6$Li atom number (markers with error bars) measured at temperatures of \{1.8, 5.8, 9.5, 16.1\}$\mu$K resepectively. The initial atom numbers for Li(Yb) are \{1.3(1.0), 1.7(3.3), 1.3(2.0), 1.8(4.0)\}$\times 10^5$ and the initial peak densities for Li(Yb) are \{6.1(2.6), 12(8.3), 10(6.2), 11(9.3)\}$\times 10^{12}$cm$^{-3}$. Fitted theoretical line shapes (solid lines) are also shown. We specify the magnetic field relative to the resonance location $B_0$ as determined from the fit. The dashed lines locate the field of maximum atom loss. The vertical axes are scaled to the theoretical background values away from the resonance. The magnetic moment of the trimer resonance state is $\delta \mu =2.8\, \mu_{\rm B}$, and $\Gamma_{\rm br}/k_B = 100$\,nK. Panel (e) shows the theoretical event rate coefficients as functions of $B-B_0$ for our four temperatures. The line colors correspond to those used in panels (a)-(d).}
	\label{fig:lsr}
\end{figure*}

The three-body recombination event rate coefficient $K_i(B,T)$ has a Lorentzian form as a function of $B$ describing the formation of a resonant trimer followed by breakup into a weakly bound dimer and a free atom. This is given by~\cite{suno2002three,wang2011numerical}
\begin{eqnarray}
K_i(B,T) &=& \left\langle (2J+1)\, 192\pi^2 \frac{\hbar}{\mu_3 k^4} |S_J(B,E)|^2 \right \rangle\,,
\label{eq:k3}
\end{eqnarray} 
where $\langle\cdots\rangle$ is the thermal average, the relative kinetic energy $E=\hbar^2k^2/(2\mu_3)$, $\mu_3$ defines the three-body reduced mass~\cite{suno2002three} and $k$ is the relative wave number. For our low temperatures, only the lowest-allowed total three-body angular momentum $J$ contributes to atom loss. We have $J=1$ for our fermion-fermion mixture. The square of the dimensionless $S$-matrix is~\cite{suno2002three,wang2011numerical,Maier2015}
\begin{eqnarray}
|S_J(B,E)|^2= \frac{\Gamma(E,J)\Gamma_{\rm br}}{(E-E_0)^2+(\Gamma(E,J)+\Gamma_{\rm br})^2/4}\,,
\end{eqnarray}
with $E_0 = \delta \mu (B-B_0)$ where $B_0$ and $\delta \mu$ are the three-body resonance location and the relative magnetic moment, respectively. A priori, $B_0$ and the two-body $B_{ \rm res}$ resonance locations need not be the same. We will conservatively assume that $B_0$ and $B_{\rm res}$ agree to within the experimental uncertainty. This uncertainty is much larger than the strength $\Delta$ of the resonance. The stimulated width is given by $\Gamma(E,J)= A (E/E_{\rm ref})^{2+J}$ with scaled width $A$ and reference energy $E_{\rm ref}$. Finally, $\Gamma_{\rm br}$ is the breakup width. The parameters $B_0$, $\delta \mu$, $A$, and $\Gamma_{\rm br}$ are determined by fitting the line shapes at different temperatures.

We now make several simplifications of Eqs.~\ref{eq:ant} consistent with our experimental system parameters. We use that the initial atom number and peak density of the two species are the same to good approximation and that the one-body loss rates satisfy $\Gamma_{\rm Li} \simeq \Gamma_{\rm Yb} = \Gamma_{\rm bg}$~\footnote{The one-body loss rates for the two species are similar to within a factor of two. This has been determined from the time evolution of atom number in single-species experiments as well as those of off-resonant two-species experiments. Three-body recombination in a spin-polarized single-species fermionic sample is negligibly small.}. Similarly, for the three-body rate we assume that $K_1(B,T) = K_2(B,T) = K(B,T)$ and $1/V_1 =1/V_2=\rho^2_{\rm Li}(\vec{x}=\vec{0})$ as both processes involve fermions that have roughly the same phase-space density and $\vec{x}=\vec{0}$ is the center of the trap. Then Eq.~\ref{eq:ant} becomes 
\begin{eqnarray}
\frac{dN_a}{dt} = -\Gamma_{\rm bg} N_a - 3 \gamma N_a^3\,,
\label{eq:td}
\end{eqnarray} 
for both $a$\,=\,Li and Yb. This differential equation has an analytic solution. 

We note that with Eq.~\ref{eq:td} we have opted for the simplest possible model that still captures the relevant physics. In particular, there exists no formal justification for our choice $K_1(B,T) = K_2(B,T)$. We are also unable to experimentally distinguish the two processes. In a recent experiment with a fermion-fermion mixture~\cite{Grimm2019}, the two processes could also not be distinguished. Reports on three-body recombination in boson-boson~\cite{baro09,Wacker2016} and boson-fermion~\cite{bloom2013tests,pires2014observation,lous2018probing} mixtures showed that the light+heavy+heavy process is much faster than the light+light+heavy one.

Figure~\ref{fig:lsr} shows atom-loss spectra and theoretical line shapes based on Eq.~\ref{eq:td} and the three-body event rate coefficient derived from Eq.~\ref{eq:k3} for a mixture with $^6$Li prepared in $|m_{s,\rm Li}, m_{i,\rm Li}\rangle = |-\nicefrac{1}{2},0 \rangle$ and $^{173}$Yb prepared in $|m_{i,\rm Yb} \rangle = |+\nicefrac{5}{2} \rangle$. Data is shown for four temperatures between 1.8\,$\mu$K and 16.1\,$\mu$K. While the three-body recombination process also causes heating of the atomic clouds, the measured temperature growth remains within $20\%$ during the first second of hold time when the three-body process is most dominant. The loss features are asymmetric and shift and broaden with increasing temperature. The parameters in the theoretical line shapes are the same for all temperatures except the one-body loss rate at our lowest temperature of 1.8\,$\mu$K, when some of our theoretical assumptions begin to break down as noted earlier. 

The satisfactory agreement between experimental data and theoretical line shapes enables us to extract the $p$-wave character of the rate coefficients. For our parameters the inequalities $\Gamma(E,J)\ll \Gamma_{\rm br} \ll k_BT$ hold and for $B>B_0$ the rate coefficient simplifies to
\begin{eqnarray}
 K(B,T) \propto \epsilon^3 e^{-\epsilon}\,,
\end{eqnarray}
with the dimensionless parameter $\epsilon$ given by ${\epsilon=\delta \mu (B-B_0)/k_BT}$. Then $K(B,T)$ has a maximum at $B=B_0 + 3k_BT/\delta \mu$, and a maximum value that is independent of temperature as shown in Fig.~\ref{fig:lsr}(e). We find that the magnetic moment of the trimer resonance is ${\delta \mu=2.8 \mu_{\rm B}}$, which can be compared to the $2\mu_{\rm B}$ magnetic moment of the diatomic $^{173}$Yb$^6$Li resonance. We expect that the trimer magnetic moment lies between 2$\mu_{\rm B}$ and 4$\mu_{\rm B}$, corresponding to a superposition state of only one $^{173}$Yb$^6$Li pair in the dimer resonant state and two $^{173}$Yb$^6$Li pairs in the resonant state. The agreement proves the $p$-wave character of our three-body recombination process and also shows that the width of the atom loss features is thermally limited even for our lowest temperature. 

We contrast our observation in this Fermi-Fermi mixture with a similar analysis of Bose gases and boson-boson mixtures which predicts that the maximum loss rate scales as $1/(k_BT)$~\cite{Maier2015}. Finally, Fig.~\ref{fig:lsr}(e) shows that the maximum three-body loss rate coefficient is relatively small for resonant processes, on the order of a few times 10$^{-27}$ cm$^6$/s, and consistent with recently reported values for the fermionic $^{40}$K and $^{162}$Dy mixture~\cite{Grimm2019}.

\section{Conclusion and outlook}

We have experimentally and theoretically studied the resonant scattering of ultracold fermionic $^6$Li and $^{173}$Yb atoms in a magnetic field. Using spin-polarized samples, we located several narrow magnetic Feshbach resonances between 580\,G and 700\,G by detecting enhanced three-body recombination near these resonances. We showed that their locations can be quantitatively explained based on the most-accurate Born-Oppenheimer potential in the literature and our own {\it ab initio} calculation of a separation-dependent hyperfine coupling between the electron spin of $^6$Li and the nuclear spin of $^{173}$Yb. 

A comparison of experimental and theoretical line profiles of the three-body recombination process at various temperatures has shown that recombination is controlled by $p$-wave scattering of the three-atom entrance channel. The observed temperature independence of the loss rate coefficient is unique to the fermionic quantum statistics of the collision partners and contrasts with the temperature dependent behavior for $s$-wave and $d$-wave bosonic scattering~\cite{Maier2015}. The analysis has also shown that the maximum recombination rate coefficient is small compared to those found for Feshbach resonances in bosonic gases.

Our observed MFRs endow the highly mass-imbalanced $^{173}$Yb-$^6$Li Fermi-Fermi mixture with strong interactions for potential applications in few- and many-body physics, and are also expected to exist in other Yb-Li isotopologues involving $^{173}$Yb or $^{171}$Yb. In particular, such resonances may aid in the pursuit of $p$-wave superfluidity in $^{173,171}$Yb-$^7$Li mixtures~\cite{cara17,scha18}. 

Our results also provide a launching pad for the production of ultracold doublet ground-state molecules. This exciting prospect may be facilitated by first producing low entropy samples of $^6$Li and $^{173}$Yb in a three-dimensional optical lattice~\cite{mose15,mark18} and then using one of the observed MFRs to coherently create YbLi molecules using magnetic field sweeps across the resonance.

\section*{Acknowledgements}

Work at University of Washington is supported by the U.S. Air Force Office of Scientific Research Grant No. FA9550-19-1-0012 and the National Science Foundation Grant No. PHY-1806212. Work at Temple University is supported by the Army Research Office Grant No. W911NF-17-1-0563 and the U.S. Air Force Office of Scientific Research Grant No. FA9550-14-1-0321. K.C.M. is supported by an IC postdoctoral fellowship.

\bibliography{mixrefs19}

\pagebreak
\onecolumngrid
\clearpage
\begin{center}
\large \textbf{Supplemental Material\\Feshbach resonances in $p$-wave three-body recombination within Fermi-Fermi mixtures of open-shell $^6$Li and closed-shell $^{173}$Yb atoms}
\end{center}
\vspace{.3in}
\twocolumngrid

\renewcommand*{\citenumfont}[1]{S#1}
\renewcommand*{\bibnumfmt}[1]{[S#1]}

\section{Spin state preparation of the atomic mixture}
\label{sec:spinprep}

To prepare a particular spin-polarized heteronuclear mixture we rely on a set of distinct manipulation and diagnostic tools for each species. 

As described in the main text, the $^6$Li sample is prepared in the $\ket{m_{s,\rm Li},m_{i,\rm Li}}=\ket{-\nicefrac{1}{2},0}$ state prior to evaporative cooling and we can subsequently transfer the spin-polarized $^6$Li sample to the hyperfine state of interest at any point in the evaporative cooling process through radiofrequency (RF) adiabatic rapid passage.

The $^6$Li spin polarization is probed by spin-dependent absorption imaging at a magnetic field of $500\,$G which relies on the fact that the electronic ground hyperfine states are separated from each other at this field by an energy that is much greater than the linewidth of the $^2{\rm S}_{\nicefrac{1}{2}}\,\rightarrow\,^2{\rm P}_{\nicefrac{3}{2}}$ probe transition. In the course of our trap-loss spectroscopy experiments, the efficiency of the RF adiabatic rapid passage is routinely checked through this spin-dependent imaging.

\begin{figure}[b]
	\includegraphics[width=1.0 \columnwidth]{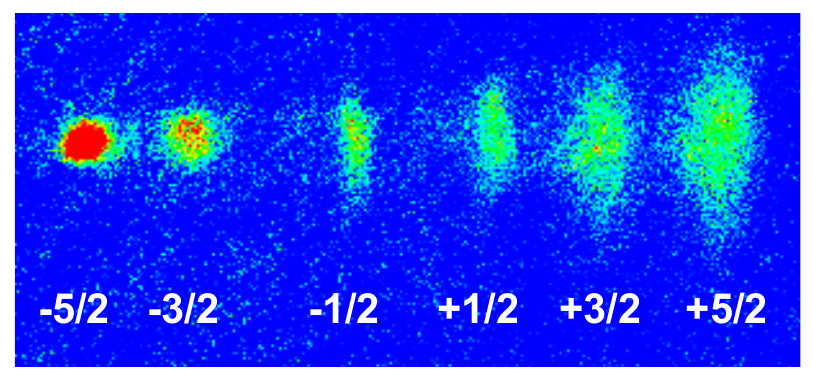}
\caption{The nuclear spin distribution of samples of $^{173}$Yb atoms. Example time-of-flight absorption image of $^{173}$Yb atoms taken after an optical Stern-Gerlach pulse (see text) is applied to an unpolarized sample with a temperature of 500$\,$nK. Each of the six clouds corresponds to $^{173}$Yb atoms in one of its six nuclear Zeeman states, labelled by $m_{i,\mathrm{Yb}}$. Prior to the optical pumping, they are roughly equally populated. No lithium atoms are present for the OSG test. The field of view is $243\,\mu{\rm m} \times 552\,\mu{\rm m}$. }
	\label{fig:osg}
\end{figure}

We prepare the desired spin polarization of $^{173}$Yb by optical pumping on the $^1{\rm S}_0 \left(f_{\rm Yb}=\nicefrac{5}{2} \right)\,\rightarrow\,^3{\rm P}_1\left(f'_{\rm Yb}=\nicefrac{7}{2}\right)$ transition. For the $^1$S$_0$ state, ${\vec f_{\rm Yb}={\vec \imath}_{\rm Yb}}$. Since the $^3$P$_1$ state has a natural linewidth of only 180 kHz, a magnetic field of 50 G is sufficient to resolve the ${f'_{\rm Yb}=\nicefrac{7}{2}}$ Zeeman states. Tuning the optical pumping laser to the electronically excited state $\ket{f'_{\rm Yb},m'_{i,\rm Yb}}=\ket{\nicefrac{7}{2},m}$ and using a combination of  $\sigma^+$ and $\sigma^-$ polarized beams pumps atoms into the $\ket{m_{i,\rm Yb}=m}$ state from the $\ket{m_{i,\rm Yb}=m \pm 1}$ states within the ground manifold. Using a sequence of such pulses, we can prepare a fully polarized sample in the targeted ground Zeeman state.

The electronic structure of the $^{173}$Yb ground state makes spin-dependent imaging infeasible. Instead, we measure the spin polarization of $^{173}$Yb samples with an optical Stern-Gerlach (OSG) method ~\cite{staie10,sstel11}, utilizing a circularly polarized laser beam which is $+860$ MHz blue detuned from the zero-field $^1{\rm S}_0 \left(f_{\rm Yb}=\nicefrac{5}{2} \right)\,\rightarrow\,^3{\rm P}_1\left(f'_{\rm Yb}=\nicefrac{7}{2}\right)$ transition frequency to create a spin-dependent force on the atoms. The OSG is performed in a $7\,$G bias field with $\sigma^+$ polarization on an atomic sample at $500\,$nK. We achieve full spatial separation between the six atomic clouds corresponding to distinct nuclear spin states in absorption images taken after time-of-flight (see Fig.~\ref{fig:osg}). In the course of our atom-loss spectroscopy, we routinely check the efficiency of the optical pumping process by performing an OSG diagnostic experiment.

We have verified that we achieve $>90\%$ spin polarization for each atomic species in the targeted spin state.

\section{Separation-dependent hyperfine interactions}
\label{sec:cpm}

We briefly expand upon our calculation of the separation-dependent hyperfine
interactions between the electron and nuclear spins of $^6$Li and the
nuclear spin of $^{173}$Yb. For $s$-wave $^{173}$Yb+$^6$Li collisions
two hyperfine coupling mechanisms are relevant.
As the two atoms move closer, electron spin density is pulled away
from the $^6$Li nucleus, which reduces the hyperfine interaction strength
between the electron and nuclear spin of $^6$Li. Simultaneously,
some of this electron spin density comes into contact with the $^{173}$Yb
nucleus and, thus, couples to its nuclear spin. These hyperfine interactions
can be written as
\begin{eqnarray}
U_{ \rm s\text{-}i }(R) = \zeta_{\rm Li}(R)\, \vec{s}_{\rm Li}\cdot\vec\imath_{\rm Li}  + \zeta_{\rm Yb}(R)\, \vec{s}_{\rm Li}\cdot \vec{\imath}_{\rm Yb}\,,
\label{eq:sitot}
\end{eqnarray}
with strengths $\zeta_{\rm Li}(R)$ and $\zeta_{\rm Yb}(R)$ that both
approach zero for large $R$. Following our definitions in the main text
the asymptotic $^6$Li hyperfine interaction $a_{\rm Li}\, \vec{s}_{\rm
Li}\cdot\vec\imath_{\rm Li}$ is accounted in the zeroth-order Hamiltonian.
The second term in Eq.~(\ref{eq:sitot}) was already introduced in the
main text.

The existence of these $R$-dependent interactions was first proposed
for RbSr~\cite{szuch10} and YbLi~\cite{sbrue12} dimers. In 2018 they
were confirmed by detecting Feshbach resonances in ultracold Rb+Sr
mixtures~\cite{sbarb18}. We will denote effects induced by the first or second
terms in Eq.~(\ref{eq:sitot}) by mechanism I and II, respectively.
Mechanism I leads to resonances where the projection $m_{f,\rm Li}=m_{s,\rm
Li}+m_{i,\rm Li}$ must be the same for the scattering and the resonant
bound state, while mechanism II leads to resonances where $m_{s,\rm
Li}+m_{i,\rm Yb}$ remains unchanged.

We have used the non-relativistic configuration-interaction valence-bond
(CI-VB) method~\cite{sKotochigova2005} to calculate the $R$-dependent
$\zeta_{\rm Li}(R)+a_{\rm Li}$ and $\zeta_{\rm Yb}(R)$ of $^{173}$Yb$^6$Li
(The CI-VB method does not compute $\zeta_{\rm Li}(R)$ directly.) The
basic idea is to construct electronic molecular wave functions from
superpositions of determinants of atomic electron orbitals localized at
the nuclear positions.  Consequently, molecular wave functions approach
a ``pure'' atomic form for large $R$, which automatically leads to
the correct molecular dissociation limits.  At small internuclear
separations orbitals around different centers have considerable overlap
and are non-orthogonal, which leads to large ``exchange effects''
that creates bonds.  We use numerical Hartree-Fock (HF) atomic
electron orbitals that avoid the need for large basis sets, as they have
the correct radial behavior near their nucleus and for large separations
between the electron and nucleus.  Furthermore, the numerical orbitals
have the correct number of nodes and are orthogonal with respect to
other HF orbitals localized at the same center.

We find that it is sufficient to use a single HF orbital for the inner
shells of Li and Yb and use only a few additional excited orbitals to
describe valence electrons. The valence orbitals are either occupied or
unoccupied HF orbitals or so-called Sturmian functions, e.g. functions
that are solutions of Hartree-Fock equations of the Coulomb problem,
where the energy is fixed and the strength of the Coulomb potential plays
the role the (generalized) eigenvalue. Sturmian functions form a complete
basis with similar asymptotic behaviour and orbital size as the occupied
valence orbitals.  Finally, our VB approach is an all-electron {\it
ab initio} calculation, which enables us to evaluate electron densities
at the nuclear sites and, thus, to calculate hyperfine structure constants
as functions of nuclear separations~\cite{sTupitsyn1998,sKotochigova1998}.

\begin{figure}
	\includegraphics[scale=0.32,trim=10 20 0 30,clip]{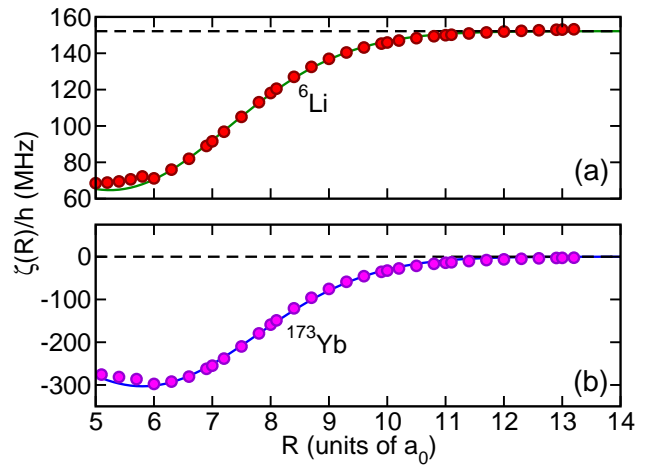}
	\caption{The hyperfine coupling constants $\zeta_{\rm
	Li}(R)+a_{\rm Li}$ (panel a) and $\zeta_{\rm Yb}(R)$ (panel b)
		as functions of separation $R$. Separations are given
		in units of the Bohr radius $a_0$ and $h$ is Planck's
		constant.  Markers represent our {\it ab~initio}
		CI-VB calculations, while the solid lines correspond to
		Gaussian fits with parameters given in 
		Table~\ref{tbl:hfspara}. Dashed lines correspond to
		the asymptotic values.}
	\label{fig:hyp}
\end{figure}

The calculated CI-VB $\zeta_{\rm Li}(R)+a_{\rm Li}$ and $\zeta_{\rm
Yb}(R)$ as functions of $R$ are shown in Fig.~\ref{fig:hyp}.  Our
asymptotic value $\zeta_{\rm Li}+a_{\rm Li}\to h\times 152.147 $ MHz is in
excellent agreement with the experimental value of $h\times 152.137$ MHz
for atomic $^6$Li~\cite{sArimondo1977}.  The strengths $\zeta_{\rm Li}(R)$
and $\zeta_{\rm Yb}(R)$ have been fit to the Gaussian $A_0 \exp(-\beta
[R-R_{c}]^2)$ and  parameters values for $A_0$, $\beta$, and $R_c$ are
given in Table~\ref{tbl:hfspara}. Small deviations between the numerical
values and the fit are noticeable for $R<6a_0$. The deviations, however, have
a negligible effect on the observable vibrationally-averaged hyperfine
strengths $\zeta_a(R)$ as the inner turning point of the most-weakly-bound
$s$-wave bound state of $^{173}$Yb$^6$Li is located at $\approx5.6a_0$.
Finally, we note that $\zeta_{\rm Li}(R)$ is three times smaller than
$\zeta_{\rm Yb}(R)$ for $R\in[5a_0,10a_0]$, where the strengths have a
significant value.

We can compare our hyperfine strengths with those of Ref.~\cite{sbrue12}
obtained using density-functional theory. Their asymptotic value of
$a_{\rm Li}$ differs by more than 5\% from the experimental value. For
completeness, the Gaussian parameters from Ref.~\cite{sbrue12} are repeated
in Table~\ref{tbl:hfspara}.

\begin{table}
	\caption{Parameters  $A_0$, $\beta$, and $R_c$ of the fitted Gaussian functions used to represent
  $\zeta_a(R)$ with $a=^6$Li or $^{173}$Yb. The last column gives the origin of the values.} 
	\begin{center}
          \begin{tabular}{cr@{$\quad$}ccc}
           \hline\hline
           $a$	&    \multicolumn{1}{c}{$A_0/h$}  & $\beta a_0^2$  & $R_c/a_0$  & Ref. \\
               &    \multicolumn{1}{c}{(MHz)}     &                &            &  \\
			\hline
      $^6$Li          & $-87.5$  &  0.1221 &  5.24 & Present\\
      $^6$Li        & $-48.8$  &  0.0535 &  4.92 & \cite{sbrue12}\\
      $^{173}$Yb      & $-303.0$ &  0.1313 &  5.81 & Present\\
      $^{173}$Yb    & $-406.0$ &  0.0868 &  6.41 & \cite{sbrue12}\\
			\hline\hline
		\end{tabular}
	\end{center}
	\label{tbl:hfspara}
\end{table}

We have performed theoretical coupled-channels calculations to locate
Feshbach resonances in the $^{173}$Yb+$^6$Li collision. The calculations
are based on the isotropic Hamiltonian using the X$^2\Sigma^+$ potential
of Ref.~\cite{sgree19}.  Table~\ref{tab:FRcomplete} gives the complete list
of predicted resonance location and strengths, ordered by magnetic field
strength and grouped by resonance mechanism.  Predictions for Feshbach
resonances near a magnetic field of 640~G have been discussed in the
main text. These resonances originate from mechanism II and have all
been observed in this work.  Two of the additional mechanism-II resonances have
been observed as shown by a comparison with the data in the table in
the main text.  Resonance strengths $\Delta$ for mechanism II are more
than one order of magnitude larger than those for mechanism I, consistent
with a larger $\zeta_a(R)$ and making them  easier to observe.

\begin{table*}[htb]
	\caption{Theoretical predictions of $s$-wave $^{173}$Yb$^6$Li
Feshbach resonances for mechanisms I and II based on coupled-channels
calculations. The first two columns describe the approximate quantum
numbers of the scattering states and bound states, respectively. The
third column is the conserved projection of the total angular
momentum. The fourth to sixth columns give the resonance position,
strength and background scattering length. Finally, the last column
identifies the resonance mechanism. The locations of resonances have a
1.3 G one-standard-deviation uncertainty due to the uncertainty of the
binding energy of the most weakly-bound state at zero applied magnetic
field~\cite{sgree19}.  The $\sim$0.1~G splittings in groups of closely-spaced
resonances are not affected by this uncertainty. }
	\label{tab:FRcomplete}
	\begin{ruledtabular}
		\begin{tabular}{ccccccc}
			\noalign{\smallskip}		
			 $|m_{s, \rm Li}^{\rm At},\ m_{i,\rm Li}^{\rm At}\rangle$ + $|m_{i,\rm Yb}^{\rm At}\rangle$ & $|m_{s, \rm Li}^{\rm Mol},\ m_{i,\rm Li}^{\rm Mol}\rangle$ + $|m_{i,\rm Yb}^{\rm Mol}\rangle$ & ${\rm M_{tot}}$  & $B_{\rm res}$ (G) & $\Delta$ ($\mu$G) & $a_{\rm bg}$ ($a_0$) & Mech.\\
			\noalign{\smallskip}
			\hline
			\noalign{\smallskip}
			 $|-1/2, 1\rangle$ + $|-3/2\rangle$ & $|1/2, 1\rangle$ + $|-5/2\rangle$   & -1 & 587.538 & 27.6 & 30.36 & \\
			 $|-1/2, 1\rangle$ + $|-1/2\rangle$ & $|1/2, 1\rangle$ + $|-3/2\rangle$   & 0 & 587.604 & 44.2 & 30.36  & \\
			 $|-1/2, 1\rangle$ + $|+1/2\rangle$ & $|1/2, 1\rangle$ + $|-1/2\rangle$   & 1 & 587.670 & 49.7 & 30.36 & II \\ 
			 $|-1/2, 1\rangle$ + $|+3/2\rangle$ & $|1/2, 1\rangle$ + $|+1/2\rangle$   & 2 & 587.737 & 44.2 & 30.36 & \\
			 $|-1/2, 1\rangle$ + $|+5/2\rangle$ & $|1/2, 1\rangle$ + $|+3/2\rangle$   & 3 & 587.803 & 27.6 & 30.36 & \\
			\\
			 $|-1/2, 1\rangle$ + $|-5/2\rangle$ & $|1/2, 0\rangle$ + $|-5/2\rangle$   & -2 & 612.284 & 2.1 & 30.36 & \\  
			 $|-1/2, 1\rangle$ + $|-3/2\rangle$ & $|1/2, 0\rangle$ + $|-3/2\rangle$   & -1 & 612.350 & 1.4 & 30.36 &  \\ 
			 $|-1/2, 1\rangle$ + $|-1/2\rangle$ &  $|1/2, 0\rangle$ + $|-1/2\rangle$   & 0 & 612.416 & 0.8 & 30.36 & I \\
			 $|-1/2, 1\rangle$ + $|+1/2\rangle$ &  $|1/2, 0\rangle$ + $|+1/2\rangle$   & 1 & 612.482 & 0.4 & 30.36 & \\
			 $|-1/2, 1\rangle$ + $|+3/2\rangle$ &  $|1/2, 0\rangle$ + $|+3/2\rangle$   & 2 & 612.549 & 0.1 & 30.36 & \\ 
			 $|-1/2, 1\rangle$ + $|+5/2\rangle$ &  $|1/2, 0\rangle$ + $|+5/2\rangle$   & 3 & & $<$ 0.1 &  & \\
			\\
			 $|-1/2, 0\rangle$ + $|-3/2\rangle$ & $|1/2, 0\rangle$ + $|-5/2\rangle$   & -2 & 639.605 & 27.6 & 30.36  & \\
			 $|-1/2, 0\rangle$ + $|-1/2\rangle$ & $|1/2, 0\rangle$ + $|-3/2\rangle$   & -1 & 639.670 & 44.2 & 30.36 & \\
			 $|-1/2, 0\rangle$ + $|+1/2\rangle$ &  $|1/2, 0\rangle$ + $|-1/2\rangle$   & 0 & 639.736 & 49.7 & 30.36 & II \\
			 $|-1/2, 0\rangle$ + $|+3/2\rangle$ &  $|1/2, 0\rangle$ + $|+1/2\rangle$   & 1 & 639.802 & 44.2 & 30.36 & \\
			 $|-1/2, 0\rangle$ + $|+5/2\rangle$ &  $|1/2, 0\rangle$ + $|+3/2\rangle$   & 2 & 639.870 & 27.6 & 30.36 & \\
			\\
			 $|-1/2, 0\rangle$ + $|-5/2\rangle$ & $|1/2, -1\rangle$ + $|-5/2\rangle$   & -3 & 666.542 & 2.3 & 30.36 & \\ 
			 $|-1/2, 0\rangle$ + $|-3/2\rangle$ & $|1/2, -1\rangle$ + $|-3/2\rangle$   & -2 & 666.608 & 1.6 & 30.36 & \\
			 $|-1/2, 0\rangle$ + $|-1/2\rangle$ & $|1/2, -1\rangle$ + $|-1/2\rangle$   & -1 & 666.674 & 0.9 & 30.36 & I \\
			 $|-1/2, 0\rangle$ + $|+1/2\rangle$ &  $|1/2, -1\rangle$ + $|+1/2\rangle$   & 0 & 666.740 & 0.5 & 30.36 & \\ 
			 $|-1/2, 0\rangle$ + $|+3/2\rangle$ &  $|1/2, -1\rangle$ + $|+3/2\rangle$   & 1 & 666.806 & 0.2 & 30.36 & \\
			 $|-1/2, 0\rangle$ + $|+5/2\rangle$ &  $|1/2, -1\rangle$ + $|+5/2\rangle$   & 2 &         & $<$ 0.1 & & \\
			\\
			 $|-1/2, -1\rangle$ + $|-3/2\rangle$ & $|1/2, -1\rangle$ + $|-5/2\rangle$   & -3 & 696.283 & 27.6 & 30.36 & \\
			 $|-1/2, -1\rangle$ + $|-1/2\rangle$ & $|1/2, -1\rangle$ + $|-3/2\rangle$   & -2 & 696.350 & 44.2 & 30.36 & \\
			  $|-1/2, -1\rangle$ + $|+1/2\rangle$ & $|1/2, -1\rangle$ + $|-1/2\rangle$   & -1 & 696.415 & 49.7 & 30.36 & II \\
			 $|-1/2, -1\rangle$ + $|+3/2\rangle$ & $|1/2, -1\rangle$ + $|+1/2\rangle$   & 0 & 696.480 & 44.2 & 30.36 & \\ 
			 $|-1/2, -1\rangle$ + $|+5/2\rangle$ & $|1/2, -1\rangle$ + $|+3/2\rangle$   & 1 & 696.545 & 27.6 & 30.36 & \\
		              \noalign{\smallskip}
		\end{tabular}
	\end{ruledtabular}
\end{table*}


%

\end{document}